\documentclass[twocolumn,amsmath,amssymb,aps]{revtex4-2}

\usepackage{gensymb} 
\usepackage{graphicx}
\usepackage{dcolumn}
\usepackage{bm}

\usepackage{enumerate} 

\begin{document}

\preprint{APS/123-QED}

\title{Quantum Transport Straintronics and Mechanical Aharonov-Bohm Effect in Quasi-metallic SWCNTs}

\author{L. Huang}

\author{G. Wei}%

\author{A.~R. Champagne}%
 \email{a.champagne@concordia.ca}
\affiliation{Department of Physics, Concordia University, Montr\'{e}al, Qu\'{e}bec, H4B 1R6, Canada}

\date{\today}%

\begin{abstract} %
Single-wall carbon nanotubes (SWCNTs) are effectively narrow ribbons of 2D materials with atomically precise edges. They are ideal systems to harness \textit{quantum transport straintronics} (QTS), i.e. using mechanical strain to control quantum transport. Their large subband energy spacing ($\sim$ 0.8 eV) leads to transistors with a single quantum transport channel. We adapt an applied model to study QTS in uniaxially-strained quasi-metallic-SWCNT transistors. The device parameters are based on an existing experimental platform, with channel lengths of $L=$ 50 nm, diameters $d\approx$ 1.5 nm, and strains up to $\varepsilon_{\text{total}}\approx$ 7 $\%$. We demonstrate that the charge carrier's propagation angle $\Theta$ is fully tunable with $\varepsilon_{\text{total}}$. When $\Theta$ reaches 90$^o$, the conductance $G$ is completely suppressed. A strain-generated band gap can be tuned up to $\approx$ 400 meV. Mechanical strain adds both scalar $\phi_{\varepsilon}$ and vector  $\textbf{A}$ gauge potentials to the transistor's Hamiltonian. These potentials create a rich spectrum of quantum interferences in $G$, which can be described as a \textit{mechanical Aharonov-Bohm effect}. The charge carriers' quantum phase can be controlled by purely mechanical means. For instance, a full 2$\pi$ phase shift can be induced in a (12,9) tube by a 0.7 $\%$ strain change. This work opens opportunities to add quantitative quantum transport strain effects to the tools box of quantum technologies based on 2D materials and their nanotubes.
\end{abstract}

\maketitle


\section{Introduction}\label{sec:intro}

Quantum transport straintronics (QTS) aims to engineer quantum-coherent charge transport using mechanical deformations (strains) in materials \cite{Fogler08, Guinea10, Amorim16, McRae19, Miao21,McRae24, Hou24}. An ideal quantum straintronics transistor should permit a complete electro-mechanical control of the electrons' Hamiltonian. For example, its charge current's quantum phase should be fully tuneable both mechanically and electrostatically. This would offers a wide range of possibilities to optimize quantum effects in 2D materials (2DMs) such as graphene, transition metal dichalcogenides (TMDs), twisted 2DMs, and 1D nanotubes/nanoribbons \cite{An24,Schock23,Shayeganfar22,Tomioka12,Minot03} made from these 2DMs. Technological applications include qubits and quantum circuits \cite{Alfieri23, Tormo_Queralt22, Mergenthaler21_2,Banszerus21, Khivrich20,Liu19}, spintronics\cite{Chen23,Pal23, Wu18, Hanakata18, Molle17, Kuemmeth08}, valleytronics \cite{Li20, Settnes17,Schaibley16,Guinea10}, superconductivity \cite{Kim23,Kapfer23, Khanjani18}, topological transitions \cite{Kim23, Zhang23, Pantaleon21, Du21, Moulsdale20, Mutch19,Efroni17}, and magnetic transitions \cite{Cenker22, Li19, Burch18}.

The experimental validation of many exciting QTS theoretical predictions \cite{Kim23,Miao21} has been a major challenge. To verify quantitatively QTS proposals, experiments require a precise control of \textit{all} sources of strain over an entire coherent-transport transistor device. This has only recently been achieved in graphene devices \cite{McRae24}. Graphene is a nearly ideal system to first bridge the existing experiment-theory divide in QTS. Its extreme mechanical strength, flexibility, and elastic deformation range make graphene's electronics vastly tunable via mechanical deformations, while its defect-free lattice supports ballistic (quantum) transport \cite{Banzerus16,Mayorov11}. Moreover, in graphene the effects of mechanical strain on charge carriers can naturally be described as a scalar potential $\phi_{\varepsilon}$, arising from changes in the next-nearest-neighbour hopping, and vector gauge potentials $\boldsymbol{A}$, from modifications to the nearest-neighbour hopping and distances \cite{CastroNeto09,Guinea10,Kitt12,Amorim16,McRae19}.

However, two experimental limitation remains to fully explore \textit{quantitative} QTS. First, graphene and 2DM crystals do not usually have atomically ordered edges and this scrambles the quantum phase of their charge carriers. One solution is to use device geometries with large width/length aspect ratios to minimize edge effects \cite{Wang21,McRae24}. Still, a large width of the transistor's channel implies that many subbands (transverse momentum modes) contribute to charge transport. Because the quantum phase of each mode is impacted differently by strain \cite{McRae24}, it makes a precise control of QTS complex. Single-wall carbon nanotubes (SWCNTs) and other nanotubes \cite{An24,Arenal07} naturally resolve these issues.

SWCNTs have perfect transverse (periodic) boundary conditions, and their very narrow width (circumference) naturally leads to a single transport mode (subband) being available at experimentally relevant charge dopings. Until now, only the most basic QTS properties of SWCNTs, such as bandgaps, have been studied \cite{Minot03, Huang08}. Thus, they represent a major unexplored opportunity for QTS experiments and models. For instance, we foresee that transistors made with a SWCNT will permit a full \textit{mechanical} control of the quantum phase of their coherent charge current, and a broad strain tuning of their electron-phonon coupling \cite{Mariani09} and spin-orbit coupling \cite{Kuemmeth08}.

In this work, we first present a comprehensive applied theoretical model that describes the ballistic conductance in uniaxially strained SWCNTs. This model is based on a previous model for graphene \cite{McRae19,Kitt12,Fogler08}, and an experimental QTS platform for low-dimensional materials \cite{McRae24}. The range of uniaxial strain we studied, $\varepsilon_{\text{total}}\leq$ 7.3\%, is experimentally feasible based on transistor dimensions of $L=$ 50 nm and tube diameters of $d\approx$ 1.5 nm. We used realistic charge doping levels for the metal-covered SWCNTs contacts \cite{McRae17,Hasegawa11,Giovannetti08,Heinze02}, $\mu_{contact}=$ 0.12 eV - 0.25 eV, and for the gate-induced Fermi level in the suspended SWCNT channel $\Delta\mu_{G}\lesssim$ 0.5 eV. Our model incorporates all of the dominant effects of uniaxial strain. The four main effects include vector potentials $\bm{A}$ resulting from alterations to the nearest-neighbor (NN) hoppings \cite{Choi10}, strain modification of the SWCNT bandgap $E_{\text{gap}}$ \cite{Yang00}, a scalar potential $\phi_{\varepsilon}$ arising from the modulation of next-nearest-neighbor (NNN) hopping \cite{Kitt12}, as well as modifications to the Fermi velocity $v_{\text{F}}$ \cite{Pellegrino11} due to strain.

We then discuss the calculated transport conductance $G$ in quasi-metallic SWCNT transistors under strain. First, we show that $G$ is tuned mechanically because $\varepsilon_{\text{total}}$ precisely controls the propagation angle, $\Theta$, and transmission, $T$, of the charge carriers across the channel. We demonstrate the ability to adjust mechanically $\Theta$ up to 90 degrees which leads to a complete suppression of $G$. A bandgap is strain-generated and widely tunable. Moreover, the strain-engineering of the quantum phase of charge carriers in the SWCNT channel leads to sharp $G$ interferences. We explain how $\varepsilon_{\text{total}}$ controls both the amplitude $\Delta G$ and phase $\Delta\Phi_{\text{FP}}$ of these interferences. We interpret this effect as a mechanical analog of the electrostatic Aharonov-Bohm effect \cite{van_Oudenaarden98, Aharonov59}. We foresee that the use of this mechanical degree of freedom to tune quantum transport, in addition to the usual electro-static ones, will expand the range of capabilities in quantum circuits and technologies.

\section{Quantum transport straintronics in SWCNT transitors} \label{Sec:model}
In this section, we will first present the conceptual QTS instrumentation and transistor geometry we used in our model. Then, we will highlight the key physical concepts and quantities needed to understand the effects of strain on quantum transport in SWCNTs. We will also explain how these are modified compared to the case of graphene \cite{McRae19}. Finally, we will summarize the derivation of the ballistic conductance $G$ as a function of experimental parameters, such as $\varepsilon_{\text{total}}$ and $V_{\text{G}}$. A more detailed derivation is presented in the Supplemental Materials (SM) \cite{SM}, and a related model for strained-graphene transistors was reported previously \cite{McRae19,McRae24}.

Figure 1(a) shows the experimental situation described by our model, where a SWCNT is freely suspended above a Si substrate (the suspension height is 150 nm) and its two ends are mechanically held by gold films (clamps). Two important points must be noted about these gold clamps. First, they are themselves suspended over an extended distance, thus forming cantilever beams. Secondly, the gold film dopes electrostatically the SWCNT sections which it covers\cite{Heinze02,McRae17} (top inset of Fig.\ 1(a)). These gold-covered sections of SWCNT form the source and drain contacts \cite{Nemec06} connecting the naked SWCNT channel in Fig.\ 1(a). The electrostatic gate voltage $V_{\text{G}}$ controlling the channel's Fermi energy, and the very small bias voltage $V_{\text{B}} \sim $ 0.1 meV used experimentally to measure $G$ are applied as shown in Fig.\ 1(a).

The three main sources of strain acting on the suspended SWCNT channel are shown in Fig.\ 1(b). In the lower inset, the electrostatic force $F_\text{G}$ \cite{McRae19,McRae24} leads to a strain $\varepsilon_\text{G}$ (blue trace) which depends on $V_{\text{G}}$ (top axis). Over the experimentally relevant range of $|V_\text{G}| \leq \pm$ 15 V, we calculate a negligibly small \cite{SM} $\varepsilon_\text{G}\lesssim$ 0.05\% due to the very short channel length. As shown in the top inset of Fig.\ 1(b), the thermal contraction (expansion) of the gold clamps (SWCNT channel) when cooling down the device to cryogenic temperatures leads to a significant, but constant, $\varepsilon_\text{thermal}$. For our device dimensions, we calculate \cite{SM} $\varepsilon_\text{thermal} \approx$ 3.2 $\%$ as shown by the gold trace. Thus, the only tunable strain in the device's channel is due to the substrate's bending as shown in Fig.\ 1(a). We label this strain component as $\varepsilon_\text{mech}$.

This mechanically tunable $\varepsilon_\text{mech}$ is along the tube's axis and applied by slightly bending the silicon substrate (ultra-thin Si) using a precision push screw. The applied strain is given by \cite{Champagne05} $\varepsilon_\text{mech} = (3ut/D^2)\delta z/L$, where $\delta z$ is the vertical displacement of the screw, $u$ is the total suspension length of the gold beams and transistor's channel, $L$ is the channel length, $t$ is the substrate thickness and $D$ is the distance between two mechanical anchoring points. We used realistic experimental values \cite{McRae24} of $L=$ 50 nm, $u = 600$ nm, $D = 8.2$ mm, $t=$ 200 $\mu$m and $\delta z$ up to $ 380 \mu$ m (limited by the failure of the Si substrate). This gives a tunable strain range of $\varepsilon_\text{mech} =$ 0 - 4.1\% as a function of $\delta z$, as shown by the black trace in Fig.\ 1(b).

Adding all sources of strain, the total strain $\varepsilon_\text{total}$ applied in our devices is tunable (approximately) from 3.2\% to 7.3\%, in experiments at low temperature. It is remarkable, and useful technologically, that $\varepsilon_\text{mech}$ is completely independent from the charge doping in the channel which is controlled via $V_\text{G}$.

When describing straintronics in SWCNTs there are some important conceptual and quantitative differences with the case of graphene. A first distinction is the transverse boundary condition (BC). The BC is periodic for SWCNTs, and it depends of how the underlying graphene layer is rolled-up to form the tube. The periodic BC for the transverse momentum $k_{y}$ can be written as $k_{y}=(2\pi/|\bm{C_\text{h}}|) (N + \alpha)$, where $N$ is an integer, and $\bm{C_\text{h}} = n\boldsymbol{a}_1 + m\boldsymbol{a}_2$. The vectors $\boldsymbol{a}_1$ and $\boldsymbol{a}_2$ are the hexagonal lattice vectors, while $n$ and $m$ are integers denoting the chirality of the tube. The length of the chirality vector $|\bm{C_\text{h}}|$ is the width $W$ of the unrolled graphene ribbon. The factor $\alpha$ is zero for the quasi-metallic tubes (i.e. ($n - m$)/3 $=$ integer \cite{Ando05}), and around 1/3 in non-metallic tubes \cite{Kolomeisky12}. As shown in Fig.\ 1(c), the consequence of this BC is the creation of a subband energy spacing $\Delta E =\hbar v_{F} (2\pi /W) \approx$ 0.8 eV when $d\approx$ 1.5 nm.

To understand transport across the transistor correctly, $\Delta E$ must be compared with $\mu_\text{contact}$. The Fermi level in the contacts is set by the charge transferred from the gold films (clamps) to the tube sections they cover. Its value ranges from $\sim$ 0.05 to 0.25 eV \cite{Heinze02} and can be controlled by Joule annealing the device \cite{McRae17, McRae24}. The inset of Fig.\ 1(c) shows the bandstructure of an unstrained quasi-metallic tube (n=14 m=8), whose $d \approx 1.5$ nm. The solid black lines indicate the allowed $k_{y}$'s set by the periodic BC. We see that in quasi-metallic tubes, the dispersion for the first subband is linear and does not have a bandgap without strain, $E_{\text{gap}}=$ 0. The dispersion of the second subband has a band gap $E_{\text{gap,2}}$, and $\Delta E = (E_{\text{gap,2}} - E_{\text{gap}})/2$. The main panel of Fig.\ 1(c) shows $\Delta E$ vs. $d$ for quasi-metallic SWCNTs.

Figure 1 (d) shows the band structure across the different sections of a quasi-metallic SWCNT transistor under strain. The source (left) and drain (right) section are unstrained and their dispersion is as shown in the inset of Fig.\ 1(c). The dashed circles indicate the position of an experimentally realistic $\mu_{\text{contact}}$ (Fermi level). We see that $\mu_{\text{contact}}$ is always much less than $\Delta E$. This is true over a broad range of SWCNT diameters (Fig.\ 1(c)). Therefore, our SWCNT transistors always have a single quantum transport channel. This is important for quantum technological applications, since a single subband wavevector $k$ means that\textit{a single quantum phase describes the entire current}. We will see below how this phase can be manipulated mechanically.

The central section of Fig.\ 1(d) shows the band structure in one valley, $K_{i}$, of a quasi-metallic channel under a strain $\varepsilon_\text{total}$. It highlights the key quantities needed to derive the ballistic conductance across the device. Using these quantities, the Hamiltonian of the charge carriers in the $K_{i}$ valley of the strained-SWCNT channel and the unstrained-SWCNT contacts are given by Eqs. \ref{Eq:H_channel} and  \ref{Eq:H_contacts}, respectively:
\begin{equation} \label{Eq:H_channel}
H_{K_{i},\text{channel}}=\hbar v_{F}\boldsymbol{\sigma}\cdot(\bar{I}+(1-\beta)\bar{\boldsymbol{\varepsilon}})
\cdot\boldsymbol{\tilde{k}}+e\phi_{\varepsilon}+\Delta\mu_{\text{G}}
\end{equation}

\begin{equation} \label{Eq:H_contacts}
H_{K_{i},\text{contact}}=\hbar v_{F}\boldsymbol{\sigma}\cdot\boldsymbol{k}+\mu_{\text{contact}}
\end{equation}
where $\bm{\tilde{k}}=\bm{k-A_{i}}$ and $\bm{k}$ are the electron's wavevectors in the channel and contacts, respectively. The Fermi velocity is $v_F = 8.8 \times 10^5$ m/s, $\bar{I}$ is the identity matrix, and $\bm{\bar{\varepsilon}}$ is the strain tensor. The parameter $\beta\approx$ 2.5 \cite{Naumis17} is the electronic Gr\"{u}neisen parameter. In the device's $x-y$ coordinates, $x$ is along the tube's axis and $y$ is along its circumference. The strain tensor $\bm{\bar{\varepsilon}}$ has elements $\varepsilon_{xx} = \varepsilon_{\mathrm{total}}, \varepsilon_{yy} = -\nu \varepsilon_{\mathrm{total}}$, and $\varepsilon_{xy} = \varepsilon_{yx} = 0$, where $\nu = 0.165$ is the Poisson ratio \cite{Naumis17}.

\begin{figure}[!htbp]
\includegraphics[scale=1]{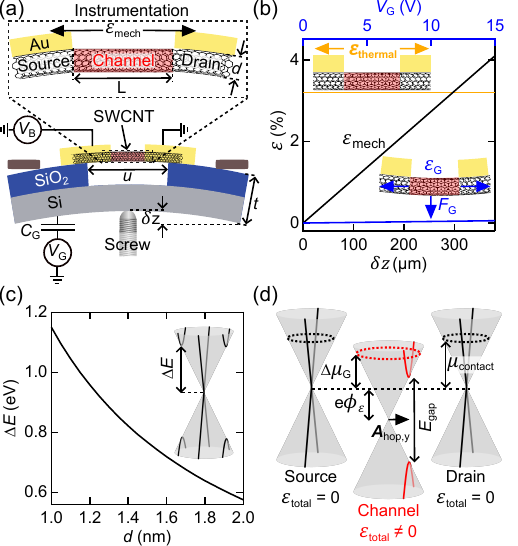}
\caption []{Electro-mechanical SWCNT quantum transistors. (a) Side-view diagram of the QTS platform we describe in our model. A push screw bends the Si substrate, which applies a $\varepsilon_\text{mech}$ to the suspended SWCNT channel. A standard dc circuit is used for transport measurements. (b) Mechanical strain (black trace), $\varepsilon_\text{mech}$, and thermal strain (gold trace), $\varepsilon_\text{thermal}$, versus $\delta z$. The gate-induced strain $\varepsilon_\text{G}$ is shown in blue as a function of $V_{\text{G}}$ (top-axis). (c) The subband energy spacing $\Delta E$ in unstrained quasi-metallic SWCNT versus their diameter $d$. The inset shows the bandstructure in one valley of an unstrained (n=14 m=8) tube. (d) Band structure in the source, channel, and drain sections of a strained quasi-metallic nanotube with $d\approx$ 1.5 nm. For clarity, the three band structures are shown as if they were disconnect (not equilibrated). We label the strain-induced scalar potential energy $e\phi_{\varepsilon}$ and vector potential component $A_{\text{hop,}y}$, the gate-induced scalar potential energy $\Delta\mu_{\text{G}}$, the strain-induced band gap $E_{\text{gap}}$, and the Fermi energy in the SWCNT contacts $\mu_{\text{contact}}$.}
\label{Fig1}
\end{figure}

The pseudospin operator $\boldsymbol{\sigma}$ is represented by the 2D Pauli matrix vector. The strain-induced vector potentials $\bm{A_{i}}$ shift the position of the Dirac (charge neutrality) points in momentum  space, and will be discussed in detail below (Fig.\ 2). The strain-induced scalar potential $\phi_{\varepsilon}$ rigidly shifts down in energy the overall band structure in the channel. Its value is \cite{Choi10,McRae24} $\phi_{\varepsilon} = g_{\varepsilon}(1-\nu)\varepsilon_{\text{total}}/e$, where $g_{\varepsilon}\approx 2.6$~eV. The quantity $\Delta\mu_G$ is the electrostatic energy induced by $V_\text{G}$ which sets the charge doping and Fermi level in the channel.

The two key strain-induced energy terms in the Hamiltonian have comparable energy scales and are shown in Fig.\ 1(d), $e\phi_{\varepsilon}$ and $\hbar v_{\text{F}}v_{\text{yy}} A_{y} = E_{\text{gap}}/2$. Where the parameter $v_{yy}=$ 1 - (1 - $\beta$)$\nu\varepsilon_{\text{total}}$ is simply the $y$-direction strain modification to the Fermi velocity. The energy shift $e\phi_\varepsilon = g_\varepsilon (1-\nu)\varepsilon_\text{total}$ reaches up to $\approx$ 0.16 eV for the maximum $\varepsilon_\text{total}$ of 7.3\%.

The magnitude of the strain-induced energies are to be compared with $\Delta\mu_G$ in the channel, see Fig.\ 1(d). This latter quantity is quantitatively different in SWCNTs compared to graphene, because the density of states $g(E)$ differs in 1D and 2D. For the first subband of a SWCNT, $g(E) =(4/\pi\hbar v_{\text{F}})(1-(E_{\text{gap}} /(2\mu_{\text{channel}}))^2)^{-1/2}$. Additionally, due to the presence of the bandgap, the so-called quantum capacitance $C_{\text{DOS}}$ \cite{Ilani06} can be quite small in SWCNTs, and thus modify significantly the total capacitance of the channel $C_{\text{total}}$. This later is the combination in series of $C_{\text{DOS}}$ and the geometric gate capacitance per unit length $C_{\text{G}} = 2\pi\epsilon_{o}/(cosh^{-1}(t_{\text{vac}}/r))$, where $t_{\text{vac}}=$ 150 nm and $r\approx$ 0.75 nm. This means that $dV_{\text{G}}= (C_{\text{DOS}}+C_{\text{G}})/(C_{\text{G}}) \times (d\mu_{\text{channel}}/e))$, see the SM for details \cite{SM}. For an experimental $V_\text{G}$ range of $\pm$ 15 V, the typically the values of $\Delta\mu_{\text{G}}$ in SWCNTs are up to $~$ 0.5 eV depending on the chirality and strain, compared to $\sim$ 0.1 eV in graphene. This enhanced ability to tune $\Delta\mu_{\text{G}}$ gives a broader spectroscopic range (tunability) to SWCNT strain-transistors.

Figure 2 describes the strain-generated vector potentials $\bm{A_{i}}$ in Eq. \ref{Eq:H_channel}, and how they open and tune bandgaps in quasi-metallic SWCNTs. Figure 2(a) shows in black ($\varepsilon_{\text{total}} =$ 0) and red ($\varepsilon_{\text{total}} =$ 7 $\%$) the first Brillouin zone (FBZ) for a quasi metallic (14,8) tube. The horizontal dashed-black and dashed-red lines indicate the positions of the allowed $k_y$ quantized values. We immediately notice that the Dirac points (charge neutrality points at $K$, $K'$ indicated by circle markers) move in $k$-space with strain. Moreover, the three-fold degeneracy of the $K$ and $K'$ valleys is lifted by strain, and we label each valley as $K_{i}$ and $K'_{i}$, with $i =$ 1, 2, 3. In Figs. 2(b)-(d), we show zoomed-in views of the  $K_1$, $K_2$, and $K_3$ valleys to better visualize the $\bm{A_{i}}$. The $\bm{A_{i}}$ are defined as the displacement between the original (black) and new (red) locations of the Dirac points of the underlying graphene lattice. By symmetry, $\bm{A'_{i}}$ in the $K'_{i}$ valleys are equal to $\bm{-A_{i}}$.

\begin{figure*}[!htbp]
\includegraphics[scale=1]{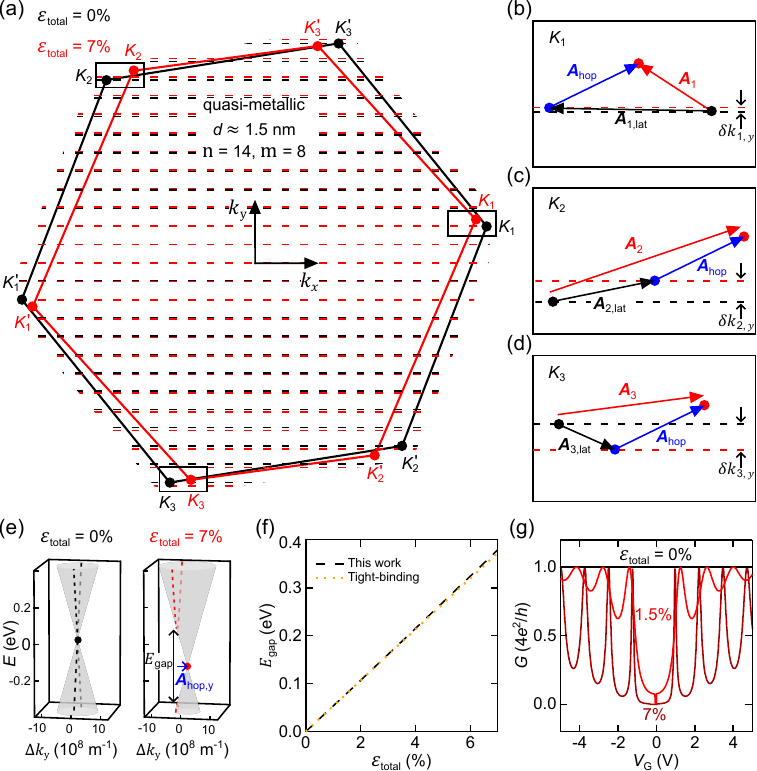}
\caption[]{Strain-tunable band structure in quasi-metallic SWCNTs. (a) The first Brillouin zones (solid lines) and sub-bands (dashed lines) for a (14,8) SWCNT are shown in black and red for $\varepsilon_{\text{total}} =$ 0 and 7 $\%$, respectively. (b), (c), (d) show zoomed-in views of the $K_1$, $K_2$, and $K_3$ valleys. (e) Energy-momentum dispersion around the Dirac points for $\varepsilon_{\text{total}} =$ 0 (left) and 7 $\%$ (right). Strain creates a band gap $E_{\text{gap}}$ due to the vector potential. (f) The calculated $E_{\text{gap}}$ vs. $\varepsilon_{\text{total}}$ for a (14,8) SWCNT in our work (dashed black) and in previous work \cite{Yang00} (dotted gold). (g) Calculated conductance $G$ versus $\text{V}_\text{G}$ at strains of 0 $\%$, 1.5 $\%$, and 7 $\%$.}
\end{figure*}

The strain modification of the NN carbon atom distances and of the electron hopping amplitudes on the hexagonal carbon lattice can be described by, $\boldsymbol{A}_{i,\text{lat}}$ and $\boldsymbol{A}_{\text{hop}}$, respectively \cite{Kitt12}. These two components and their sum, $\bm{A_{i}}$, are shown in Figs.\ 2(b)-(d). Equation \ref{Eq:A} gives the expression to calculate these vector potentials in SWCNTs.

\begin{subequations} \label{Eq:A}
\begin{equation}
\begin{aligned}
\boldsymbol{A}_{\text{hop}}& =\frac{\beta\varepsilon(1+\nu)}{2a}\begin{pmatrix}\sin3\theta_h\\\\\cos3\theta_h\end{pmatrix}  \\
\end{aligned}
\end{equation}
\\
\begin{equation}
\begin{aligned}
\boldsymbol{A}_{1,\text{lat}}& =\frac{2\pi\varepsilon}{3a}\begin{pmatrix}-\cos\theta_h - \frac{1}{\sqrt{3}}\sin\theta_h\\\\\frac{1}{\sqrt{3}}\nu\cos\theta_h -\nu\sin\theta_h\end{pmatrix}
\end{aligned}
\end{equation}
\\
\begin{equation}
\begin{aligned}
\boldsymbol{A}_{2,\text{lat}}& =\frac{2\pi\varepsilon}{3a}\begin{pmatrix}\cos\theta_h-\frac{1}{\sqrt{3}}\sin\theta_h\\\\\frac{1}{\sqrt{3}}\nu\cos\theta_h + \nu\sin\theta_h\end{pmatrix}
\end{aligned}
\end{equation}
\\
\begin{equation}
\begin{aligned}
\boldsymbol{A}_{3,\text{lat}}& =\frac{4\pi\varepsilon}{3\sqrt{3}a}\begin{pmatrix}\sin\theta_h\\\\-\nu\cos\theta_h\end{pmatrix}
\end{aligned}
\end{equation}
\end{subequations}
Here, $a=$1.42 $\text{Å}$ is the NN carbon-carbon distance without strain. $\theta_h$ is the angle between the chiral vector $\boldsymbol{C}_{h}$ and the zigzag lattice direction (Fig.\ S2a) \cite{Charlier07,SM}. We note that the definition of $\theta_h$ differs from the definition of the lattice orientation used in graphene devices \cite{McRae24}.

Stretching tubes longitudinally also decreases their circumference due to the Poisson ratio, i.e. $\varepsilon_{yy} = -\nu \varepsilon_{\mathrm{total}}$. Because the subband energy spacing $\Delta E$ depends on the circumference, a special attention must be paid to this change. Under strain, the periodic BC along the circumference becomes:
\begin{equation}
\tilde{k_y}(1+\varepsilon_{yy})C_h=2\pi N
\end{equation}
\\
where $N$ is an integer. As can be seen in Figs.\ 2(a)-(d), in the absence of strain the dashed-black subbands pass through the black Dirac points in quasi-metallic nanotubes. This translates to a zero bandgap as shown in Fig.\ 2(e). When strain is applied, the diameter of the tubes shrinks and $\Delta E$ increases. The strained (dashed-red) subbands in Figs.\ 2(a)-(d) no longer pass through the red Dirac points. A close inspection of Figs.\ 2(b)-(d) reveals that the change in $\Delta E$ shifts the allowed $k_{y}$ values. The $k_{y}$ shifts are exactly equal to $A_{i,lat,y}$, and thus always cancel them out. Meaning that the allowed $y$-component of the generalized wave vector in the strained channel are $(k_y + A_{i,\text{lat},y}) - (A_{i,\text{lat},y} + A_{\text{hop},y}) = k_y - A_{\text{hop},y}$. Thus, only the $A_{\text{hop},y}$ part of the vector potentials matters in determining the allowed electron states.

Figure 2(f) shows that the $A_{\text{hop},y}$ also lead to strain-tunable band gaps. The calculated values for $E_\text{gap}=2\hbar v_{\text{F}}v_{yy}A_{\text{hop},y}$ as a function of $\varepsilon_{\text{total}}$ are shown in Fig.\ 2(g), and are in complete agreement with previous work \cite{Yang00}.

The next step towards our objective of calculating $G$ across the transistors is to solve for the transmission amplitude $t_{\xi}$ of the carriers. We do this by using the longitudinal BCs for the wavefunction at the source-channel and channel-drain interfaces in each valleys $\xi=$ $\pm$1, where the $+$ and $-$ refer to $K_{i}$ and $K'_{i}$, respectively. This step is detailed in the SM \cite{SM}, and in previous works \cite{Tworzydlo06,McRae19,McRae24}. The transmission probability is $T_{\xi} = |t_{\xi}|^2$, and its expression is given in Eq. \ref{Eq:T_n_Str_A}.
\begin{widetext}
\begin{equation} \label{Eq:T_n_Str_A}
T_{\xi}=\frac{(k_x\tilde{k_x}v_{xx})^2}{(k_x\tilde{k_x}v_{xx})^2\cos^2(\tilde{k_x}L)+
(k_F\tilde{k_F}-k_y(k_y-\xi A_{\text{hop},y})v_{yy})^2\sin^2(\tilde{k_x}L)}
\end{equation}
\end{widetext}
where $k_F = \mu_{\text{cont}}/(\hbar v_{\text{F}})$ and $\tilde{k_\text{F}} = \mu_{\text{channel}}/(\hbar v_{\text{F}})$. The strain modified Fermi velocity factor along $x$ and $y$ are $v_{xx}=1+(1-\beta)\varepsilon_{\text{total}}$ and $v_{yy}=1-(1-\beta)\nu\varepsilon_{\text{total}}$. Therefore, $k_x = \sqrt{{k_F}^2-{k_y}^2}$, and $\tilde{k_x}=v_{xx}^{-1}\sqrt{\tilde{k_F}^2-(v_{yy}\tilde{k_y})^2}=v_{xx}^{-1}\sqrt{\tilde{k_F}^2-({k_y - \xi A_{\text{hop},y}})^{2}v_{yy}^2}$.

Finally, the ballistic conductance is
\begin{equation}\label{Eq:G}
G=\frac{4e^2}{h}\frac{1}{2}\sum_\xi T_{\xi}
\end{equation}

We used this expression to calculate the $G - V_{\text{G}}$ data plotted in Fig.\ 2(h) for a quasi-metallic (14,8) device with strain $\varepsilon_{\text{total}}=$  0 (black), 1.5 $\%$ (pale red), and 7 $\%$ (dark red). The effect of the strain-induced $\phi_{\varepsilon}$ is to shift to lower $V_{G}$ the $G$ minimum as strain increases. A bandgap is clearly visible in the $\varepsilon_{\text{total}}=$ 7 $\%$ data. Beyond the complete suppression of $G$ inside the bandgap, we observe that $A_{\text{i,y}}$ reduces $G$ at all $V_{G}$'s and creates interferences. We stress that these Fabry-P\'{e}rot interferences take place in a single quantum transport channel. The next section discusses how they can be precisely strain-engineered \textit{in-situ}.

\section{Mechanical Aharonov-Bohm effect in quasi-metallic SWCNTs}

Quantum transport circuits \cite{Alfieri23,Pal23,Tormo_Queralt22} are usually manipulated via multiple electric and magnetic fields, but their mechanical degrees of freedom mostly remains unused and poorly controlled. A precise mechanical control of quantum transport phases would add both new functionalities and reduce quantum dephasing, due to mechanical disorder, in devices such as quantum bits, transmission lines, transistors, and NEMS.

Figure 3(a) shows $G-\Delta \mu_{\text{G}}-\varepsilon_{total}$ data calculated using Eq.\ \ref{Eq:G} for a (14,8) quasi-metallic tube (similar data for a (12,9) tube are shown in Fig.\ 4). We readily notice four qualitative features. First, a band gap opens up as strain increases (blue region). Secondly, the center of this $G=$ 0 region (off state) is shifting linearly downward with strain due to $e\phi_{\varepsilon}$. Thirdly, even outside the bandgap region, $G$ is decreasing as $\varepsilon_{total}$ increases (see also Fig.\ 2(g)). Finally, we observe sharp $G$ interferences whose energy spacing is $\Delta E_{\text{FP}} \approx hv_{\text{F}}/2L\approx$ 36 meV, where $L =$ 50 nm is the length of the channel. The positions of these interferences shift with strain. We first seek to describe quantitatively the strain-dependence of these Fabry-P\'{e}rot (FP) interferences in terms of electron trajectories as a function of $\varepsilon_{total}$.

Figure 3(b) shows the charge carriers' trajectories in quasi-metallic SWCNTs, in the first and only occupied subband, when $\varepsilon_{\text{total}}=$ 0 (top diagram) and $\neq$ 0 (bottom diagram). As discussed above, the effect of strain is two fold. First, $e\phi_{\varepsilon}$ modifies the magnitude of the Fermi wavevector in the channel. For example, when $V_{\text{G}}>$ 0 then $e\phi_{\varepsilon}$ increases $\mu_{\text{channel}}$ and $|\bm{\tilde{k}|}$. Secondly, $A_{i,y}=\pm A_{hop,y}$ modifies the orientation of $\bm{\tilde{k}}$. The current flows at a propagation angle $\pm\Theta = sin^{-1}(A_{hop,y}/|\bm{\tilde{k}|})$ with respect to the tube's axis.

It is well known \cite{CastroNeto09}, and consistent with Eq.\ \ref{Eq:T_n_Str_A}, that $G$ strongly depends on $\Theta$ in SWCNTs and graphene. What is unique about the present devices is that a purely mechanical degree of freedom, uniaxial-strain, permits a precise control of $\Theta$ over its entire range from 0$^o$ to 90$^o$. Figure 3(c) shows the transmission probability, $T$, and $G= 4e^2/h T$ vs $\Theta$ in a quasi-metallic (14,8) transistor. Clear oscillations (interferences) of $G$ vs. $\Theta$ are visible. The relationship between $\Theta$ and $\varepsilon_{total}$ is shown in Fig.\ 3(d).

The black data in Fig.\ 3(d) is extracted along the horizontal black cut in Fig.\ 3(a) at $\mu_{\text{channel}}=$ -0.2 eV. The transmission angle $\Theta$ evolves smoothly from 0$^o$ to 90$^o$ as $\varepsilon_{total}$ changes from 0 to $\approx$ 4 $\%$. The $\Theta=$90$^o$ point corresponds to crossing into the band gap region in Fig.\ 3(a). For completeness, we show in Fig.\ 3(d) that a similar control of $\Theta$ is possible in other types of SWCNTs. The dotted data are for a semiconducting (16,5) tube. The only instance when $\Theta$ is not strain-tunable is in armchair tubes, as shown by the dashed (11,11) data. Our prediction of a precise mechanical control of $\Theta$ and $G$ in quasi-metallic SWCNT transistors is robust over a broad range of tube diameters beyond $d=$ 1.5 nm. For instance, we show in the SM \cite{SM} that the picture is unchanged for 1.0 nm $<d<$ 2.0 nm. The strain-dependence of the FP interferences in Fig.\ 3(a) also implies a mechanical control of the quantum phase of charge carriers. We now discuss this phase control and how it could be used in devices.

\begin{figure}[!htbp]
\includegraphics[scale=1]{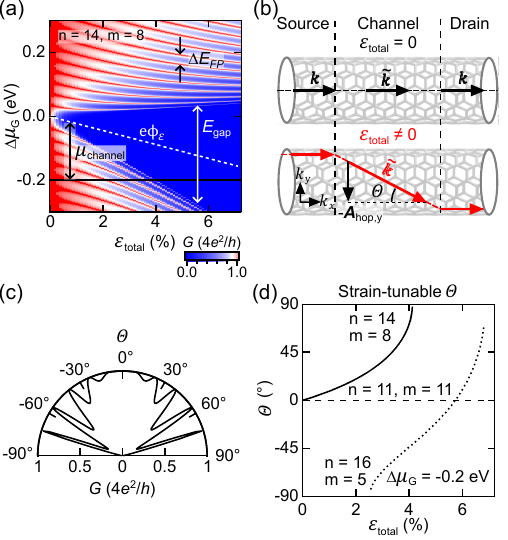}
\caption[]{Mechanical control of the propagation angle and transmission probability of charge carriers. (a) $G$-$\Delta \mu_{\text{G}}$-$\varepsilon_{\text{total}}$ data calculated for a (14,8) quasi-metallic tube. Strain opens a band gap, as shown by the blue region. A linear shift is created by the scalar potential energy $e\phi_{\varepsilon}$ (white dashed line). Clear FP resonances, with spacing $\Delta E_{\text{FP}}$, are visible. (b) Trajectory of charge carriers in the first subband of quasi-metallic nanotubes under zero (black line) and finite (red line) strain. The charge carrier propagation angle $\Theta$ is strain tunable. (c) Polar plot of $G$ versus $\Theta$ for a (14,8) tube when $\mu_{\text{G}} = -0.2$ eV, i.e. along the black cut in panel (a). (d) Strain controls $\Theta$ in all quasi-metallic (e.g. $n = 14,\ m = 8$ data shown as a solid line) and semi-conducting (e.g. $n = 16,\ m = 5$ data shown as a dotted line) SWCNTs. However, $\Theta$ is independent from $\varepsilon_{total}$ in armchair tubes (e.g. $n = 11,\ m = 11$ data shown as a dashed line).}
\end{figure}

Figure 4(a) is a diagram representing the two most likely paths followed by ballistic charge carriers across a quasi-metallic SWCNT channel under strain. Path $\#$1 is a single passage across the channel at a transmission angle $\Theta = sin^{-1}(A_{hop,y}/|\tilde{\bm{k}}|)$. Path $\#$2 is for charge carriers which are reflected upon first encountering the channel-drain interface, but transmitted into the drain contact upon their second attempt. Higher order terms (paths) which involves more reflections, and make smaller contributions to $G$, are not shown for clarity.

The measured transmission probability $T$, and conductance $G$, is obtained by squaring the sum of the amplitudes of these superposed paths. Therefore, $G$ will oscillate as a function of the additionnal Fabry-P\'{e}rot phase $\Phi_{\text{FP}}$ acquired along path $\#2$. That $\Phi_{\text{FP}}$ and interference amplitude $\Delta G$ can be tuned electrostatically is a well established result \cite{Liang01,Lotfizadeh21}. What is new, and technologically useful, is that $\Phi_{\text{FP}}$ and $\Delta G$ can also be tuned entirely mechanically.

Figure 4(b) shows the $G - \Delta \mu_{\text{G}}-\varepsilon_{total}$ for a (12,9) quasi-metallic tube (similar data for a (14,8) tube are shown in Fig.\ 3). The oblique dashed lines in the bottom left of the panel show data cuts at $\Theta =$ 10, 30, and 50$^o$. Along these constant angle cuts it is easier to calculate $\Phi_{FP}= \oint \bm{\tilde{k}}\cdot\bm{dl}$ where $\bm{dl}$ is the displacement around the shaded-red triangular loop in Fig.\ 4(a). Note that $\bm{A}$ is constant everywhere in the channel, and only the $\tilde{k_{x}}$ component is inverted when the wavefunction is reflected. We thus find $\Phi_{\text{FP}} =$ 2$\tilde{k_{x}}L$. This is fully consistent with Eq.\ \ref{Eq:T_n_Str_A} whose phase dependence origins from the functions $sin^{2}(\tilde{k_{x}}L)$ and $cos^{2}(\tilde{k_{x}}L)$.

Using the model in Section \ref{Sec:model}, we can re-write $\tilde{k_{x}}$ to better visualize the strain dependence of $\Phi_{\text{FP}}$ as
\begin{equation}\label{Eq:FP}
\Phi_{\text{FP}} = 2 \sqrt{[(\Delta \mu_{\text{G}}+e\phi_{\varepsilon})/(\hbar v_{\text{F}})]^{2} - (A_{\text{hop},y}v_{yy})^2}(\frac{L}{v_{xx}})
\end{equation}

Equation \ref{Eq:FP} highlights that $\Phi_{FP}$ can be controlled via $\phi_{\varepsilon}$ and $A_{\text{hop},y}$, which are both proportional to $\varepsilon_{\text{total}}$. This mechanical control of quantum interferences is closely analogous to the usual \textit{electrostatic} Aharonov-Bohm (AB) phase, where an electrostatic potential can be used to tune $G$ interferences \cite{Aharonov59,van_Oudenaarden98}. We can verify this analogy by comparing the mechanical tuning of the phase, $\Phi_{\text{FP,mech}}$, and the electrostatically-induced phase $\Phi_{\text{FP,electro}}$ controlled via $V_{\text{G}}$.

Figure 4(c) shows $G - \Phi_{\text{FP,electro}}$ data extracted along the vertical dotted line in Fig.\ 4(b). The only parameter tuning the FP phase along this cut is the electrostatic potential $V_{\text{G}}$. For instance, $\Phi_{\text{FP,electro}}$ can be changed by $\pi$, and $G$ from a local minimum to a maximum, between $V_{\text{G}}=$ 1.29 V and 2.03 V. Many full oscillations of $\Phi_{\text{FP,electro}}$ are possible over the experimentally available $V_{\text{G}}$ range of $\pm$ 15 V.

In Fig.\ 4(d), we show data extracted along the tilted dashed lines, in Fig.\ 4(b), for which $\Theta =$ 10$^o$, 30$^o$, and 50$^o$. Many complete $G$ oscillations are visible along these cuts. To move along these constant $\Theta$  cuts, and tune $\Phi_{\text{FP,electro-mech}}$ and $\Delta G$, one must adjust both the electrostatic, $V_{\text{G}}$, and mechanical potentials, $\phi_{\varepsilon}$, $A_{\text{hop},y}$.

Lastly, in Fig.\ 4(e) we show $G$ interferences which are controlled solely via the mechanically-generated potentials. These data cuts are extracted parallel to the $\varepsilon_{total}$ axis in Fig.\ 4(b) and show many complete oscillations of $G$. For instance, in the black trace a full 2$\pi$ change in $\Phi_{\text{FP,mech}}$ is created by changing $\varepsilon_{\text{total}}$ from 4.4 $\%$ to 5.1$\%$. The amplitude of the interferences $\Delta G$ can be engineered with strain and can become $\sim$ 4$e^2/h$. The data in Fig.\ 4(e) constitute a mechanical analog to the electrostatic Aharonov-Bohm effect.

Collectively, the data in Fig.\ 4 make a clear prediction for realistic experiments: quasi-metallic SWCNT ballistic transistors can serve as coherent current sources whose phase is mechanically tunable. This additional degree of freedom to control the phase of current in quantum circuits, independently from electrostatics, opens new possibilities for previous work on SWCNT qubits and coherent transport devices \cite{Alfieri23,Chen23,Tormo_Queralt22}.

\begin{figure}[!htbp]
\includegraphics[scale=1]{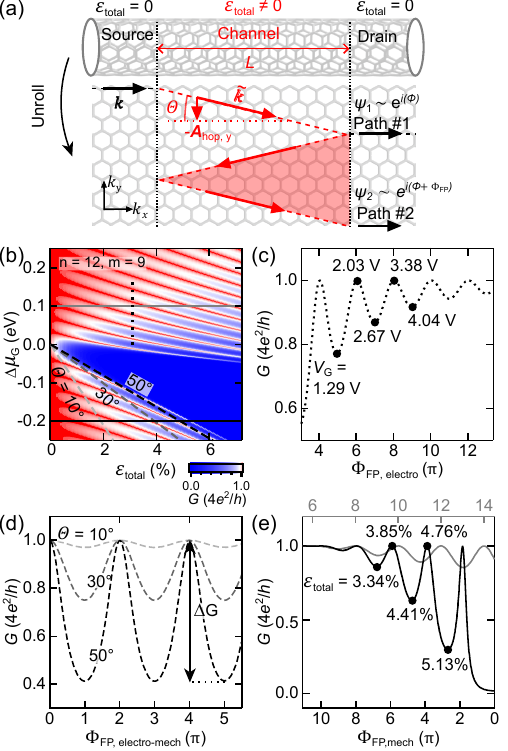}
\caption[]{Mechanical control of the quantum transport phase in nanotube transistors. (a) Diagram of the charge carriers trajectories across a quasi-metallic SWCNT channel under strain. Paths $\#$1 and $\#$2 interfere with a phase difference $\Phi_{\text{FP}}$ which depends on $\Theta$ and is mechanically tunable. (b) Calculated $G - \Delta \mu_{\text{G}} - \varepsilon_{\text{total}}$ data for a (12,9)-SWCNT transistor. (c) Data extracted along the vertical dotted line in panel (b). These $G$ interferences are tunable electrostatically using $V_{\text{G}}$. (d) Data extracted along the tilted dashed lines in panel (b) where $\Theta = $ 10$^o$, 30$^o$, and 50$^o$. The $G$ oscillations are tunable using a combination of electrostatic, $V_{\text{G}}$, and mechanical potentials $\phi_{\varepsilon}$, $A_{hop,y}$. (e) Data extracted along the horizontal solid lines in panel (b). The $G$ interferences are fully tunable using only the mechanical potentials.}
\end{figure}

\section{Conclusions}

In conclusion, single-wall nanotubes are very narrow ribbons of 2D materials with atomically precise edges, making them ideal systems for QTS (quantum transport straintronics, i.e. using mechanical strain to control quantum transport). Their narrow widths means that it is realistic to build transistors whose transport involves a single subband, and where a single quantum phase describes the conductance or current. Here, we developed an applied theory, describing an existing experimental platform \cite{McRae24}, to explore QTS and a mechanical Aharonov-Bohm effect in quasi-metallic SWCNTs.

A uniaxial mechanical strain $\varepsilon_{\text{total}}$ in a SWCNT creates a scalar, $\phi_{\varepsilon}$, and vector, $\boldsymbol{A} = \boldsymbol{A}_{\text{lat},i} + \boldsymbol{A}_{\text{hop}}$, gauge potentials. The former is analogous to a mechanical gating effect or tuning of the work function. Its experimental signature differs quantitatively from the one in graphene due to the differences in densities of states and quantum capacitances. The $\boldsymbol{A}$ potentials acting on the charge carriers are also distinct from the ones in graphene because of the periodic boundary condition in SWCNTs. The strain-dependent diameter of a tube generates a momentum-shift of its subbands, which cancels the $\boldsymbol{A}_{\text{lat}}$ component present in graphene \cite{McRae19}. This has an important consequence, charge carriers in all valleys of the SWCNT Hamiltonian have the same $|\boldsymbol{A}|$ and the same propagation angle $|\Theta|$. Thus, all charge carriers propagating across the transistor are coherent.

We showed that a realistic experimental range of $\varepsilon_{\text{total}} \approx$ 0 - 7 $\%$ is more than sufficient to tune $\Theta$ from 0$^o$ to 90$^o$, in a broad range of quasi-metallic SWCNTs. When $\Theta =$ 90$^o$, $G$ is reduced to zero as the channel's Fermi energy enters the band gap. The numerical value of the strain-tunable bandgap is in agreement with previous semi-classical calculations \cite{Yang00}. The calculated transport data $G - \Delta \mu_{\text{G}} - \varepsilon_{\text{total}} $ shows a rich set of quantum interferences which were not previously reported and discussed.

We explained this interference spectrum in terms of a phase $\Phi_{FP}$ added by the mechanically-induced potentials as well as by the electrostatic (gate) potential. In quasi-metallic SWCNTs, it is possible to tune $\Phi_{FP}$ with $\varepsilon_{\text{total}}$ over several 2$\pi$ oscillations, as well as control the amplitude $\Delta G$ of the interferences. This effect can be interpreted as a mechanical analog of the electrostatic Aharonov-Bohm effect \cite{Aharonov59,van_Oudenaarden98}.

This work opens many opportunities to explore and harness quantitative strain effects in quantum technologies \cite{Alfieri23, Pal23, Chen23, Tormo_Queralt22, Mergenthaler21,Khivrich20} and condensed matter physics \cite{Szombathy24,Zhang23,Lotfizadeh21,Anderson21}. A detailed understanding of strain effects on quantum transport will also be useful to optimize existing transport devices based on 2DMs and their nanotubes, which are invariably subjected to strains inside heterostructures \cite{Hou24,Kogl23}. We believe that experiment-ready applied theoretical models, such as the one presented here, can help accelerate the development of QTS in 1D and 2D materials.
\\

\section{Data Availability Statement}
The data that support the findings of this study are included in the Figs. of the main text and Supplementary Material. They are also available from the corresponding author upon reasonable request.

\begin{acknowledgments}
This work was supported by NSERC (Canada), CFI (Canada), and Concordia University. We acknowledge usage of the QNI (Quebec Nano Infrastructure) cleanroom network.
\end{acknowledgments}

\appendix

\bibliography{main}

\end{document}


\maketitle
\tableofcontents

\section{Thermal and electrostatic strain in the studied \\ SWCNT transistors}\label{sec:S1}
As shown in Figs.\ 1(a)-(b) of the main text, the proposed SWCNT transistors are subject to three sources of strain: thermal, electrostatic, and mechanical. The mechanical strain is generated by bending the substrate on which the sample is fabricated. The thermally-induced strain arises during the cooling of the device from room temperature to cryogenic temperatures. The electrostatic strain is caused by the electrostatic force between the gate and the suspended channel. In this section, we calculate realistic estimates of the thermal strain and the electrostatic strain based on the device parameters presented in the main text.

\subsection{Thermal strain}
The thermal strain $\varepsilon_\text{thermal}$ in the channel has two main sources: the thermal contraction of the Au cantilever beams $\varepsilon_\text{th,Au}$ and the thermal expansion of the carbon nanotube channel $\varepsilon_\text{th,CNT}$. The impact from the contraction of the Si/SiO$_2$ substrate is negligible\cite{McRae19}. $\varepsilon_\text{thermal}$ is given by:

\begin{equation}
\varepsilon_{\text{thermal}}= \varepsilon_{\text{th,SWCNT}}+\varepsilon_{\text{th,Au}}
=\int_{295 K}^{ 1 K}\alpha_{\mathrm{SWCNT}}(T)\text{d}T-\frac{u-L}{L}\int_{295 K}^{1 K}\alpha_{\mathrm{Au}}(T)\text{d}T
\end{equation}
\\
where $\alpha_{\text{SWCNT}}$ \cite{Yoon11} and $\alpha_{\text{Au}}$ \cite{Nix41} are the expansion coefficients of graphene and Au, respectively. The total suspension length $u = 600$ nm is the sum of the suspended cantilever beam lengths plus the channel length, and the channel length $L = 50$ nm. At $T = 1$ K, $\varepsilon_{\text{thermal}} \approx 3.2\%$.

\subsection{Electrostatic strain}
The electrostatic strain $\varepsilon_\text{G}$ is caused by the electrostatic force $F_{G}$ between the gate and the nanotube channel. An electrostatic force also exists between the gate and the Au cantilevers. However, given the 50 nm thickness of the Au cantilevers, their bending is completely negligible\cite{Island12}.

$\varepsilon_{\text{thermal}} \approx 3.2\%$ ensures that there is no slack in the SWCNT channel independently of the gate voltage applied. Thus, the carbon nanotube remains in the elastic stretching regime. One can then calculate the electrostatic force and the tension as \cite{Sazonova06}:

\begin{equation}
F_{\mathrm{G}}=\frac12C_{\mathrm{G}}'V_{\mathrm{G}}^2
\end{equation}

\begin{equation}
T_\text{G}=\left(\frac{Y S}{8}\right)^{\frac{1}{3}}F_{\mathrm{G}}^{\frac{2}{3}}
\end{equation}
\\
where $C_\text{G}' = \text{d}C_\text{G}/\text{d}t_\text{vac}$ is the derivative of the gate capacitance with respect to the vacuum spacing between the gate and the channel $t_\text{vac}$, $Y = 1$ TPa is Young's modulus of a SWCNT, $S$ is the tube cross-sectional area. The capacitance can be modelled as a wire over a plane \cite{McRae17}:

\begin{equation} \label{Eq:C_G}
C_\text{G}=\frac{2\pi\varepsilon_\text{vac}}{\cosh^{-1}(2t_\text{vac}/d)}L
\end{equation}
\\
where $\varepsilon_\text{vac}$ is the vacuum permittivity, $d$ is the tube diameter and $L$ is the channel length. In our proposed SWCNT devices: $t_\text{vac} = 150$ nm, $d = 1.5$ nm, $L = 50$ nm, thus, $C_\text{G} \approx 4.64 \times 10^{-19}$ F, and $C_\text{G}' \approx 5.16 \times 10^{-13}$ F/m. Using the maximum experimentally realistic gate voltage (in a suspended device) of 15 V, $F_\text{G} \approx 5.81 \times 10^{-11}$ N, $T_\text{G} \approx 8.73 \times 10^{-10}$ N, $\varepsilon_\text{G} = T_\text{G}/(Y S) \approx 0.06$ \%. Comparing to the $\varepsilon_{\text{thermal}} \approx 3.2\%$, $\varepsilon_\text{G}$ is completely negligible at all values of gate voltage.

\section{Derivation of the ballistic conductance in strained SWCNT transistors}
In this section, we derive step-by-step the applied theoretical model we used to calculate the conductance in strained SWCNT transistors. We begin by incorporating all realistic experimental parameters and strain-induced effects into the Hamiltonian. Then, we solve the Dirac equation (graphene) with uniaxial strain to obtain the plane wave solutions. Next, we use the boundary conditions at the source-channel-drain interfaces to determine the expression for the transmission amplitude. We impose the periodic boundary conditions in the transverse ($y$) direction appropriate for SWCNTs. It allows us to calculate the strain-induced shifts of the Dirac points and sub-band positions in the SWCNT-channel's First-Brillouin zone. We then identify which subband contributes to charge transport and calculate the value of $G$.

\subsection{Hamiltonian and eigenstates in strained SWCNTs}
The Hamiltonian in the strained-SWCNT channel and unstrained-SWCNT contacts are given by Eq. \ref{Eq:H_channel} and Eq. \ref{Eq:H_contacts}:

\begin{equation} \label{Eq:H_channel}
H_{\text{channel}}=\hbar v_{F}\boldsymbol{\sigma}\cdot(\bar{I}+(1-\beta)\bar{\boldsymbol{\varepsilon}})\cdot\boldsymbol{\tilde{k}}+\Delta\mu_{\text{G}}+e\phi_{\varepsilon}
\end{equation}

\begin{equation} \label{Eq:H_contacts}
H_{\text{contact}}=\hbar v_{F}\boldsymbol{\sigma}\cdot\boldsymbol{k}+\mu_{\text{contact}}
\end{equation}
\\
where $\bm{\tilde{k}}=\bm{k-A_{i}}$ and $\bm{k}$ are the electron's wavevectors in the channel and contacts, respectively. The Fermi velocity is $v_F = 8.8 \times 10^5$ m/s, $\bar{I}$ is the identity matrix, $\bm{\bar{\varepsilon}}$ is the strain tensor. The parameter $\beta\approx$ 2.5 \cite{Naumis17} is the electronic Gr\"{u}neisen parameter. In the device's $x-y$ coordinates, $x$ is along the tube's axis and $y$ is along its circumference. The strain tensor $\bm{\bar{\varepsilon}}$ has elements $\varepsilon_{xx} = \varepsilon_{\mathrm{total}}, \varepsilon_{yy} = -\nu \varepsilon_{\mathrm{total}}$, and $\varepsilon_{xy} = \varepsilon_{yx} = 0$, where $\nu = 0.165$ is the Poisson ratio \cite{Naumis17}. The pseudospin operator $\boldsymbol{\sigma} = (\sigma_x, \sigma_y)$ is represented by the Pauli matrices and acts on the two-component spinor wavefunction associated with the A and B sublattices. The pseudospin orientation aligns either parallel (up) or anti-parallel (down) with the generalized wave vector $\boldsymbol{\tilde{k}}$. The term $\Delta\mu_G$ represents the gate-induced electrostatic potential.

Uniaxial strain induces four main qualitative effects on the channel’s band structure. First, it causes a downward shift of the Fermi energy, which can be described by a scalar potential $e\phi_{\varepsilon} = g_{\varepsilon}(1 - \nu)\varepsilon_{\mathrm{total}}$, where $g_{\varepsilon} \approx 2.6$ \text{eV} \cite{Choi10, McRae24}. Second, it leads to an anisotropic distortion of the Dirac cones, resulting in a direction-dependent Fermi velocity $\boldsymbol{\bar{v}_F} = v_F(\boldsymbol{\bar{I}} + (1 - \beta)\bar{\boldsymbol{\varepsilon}})$. For uniaxial strain, the matrix $(\boldsymbol{\bar{I}} + (1 - \beta)\bar{\boldsymbol{\varepsilon}})$ only has diagonal elements $v_{xx} = 1 + (1 - \beta)\varepsilon_{\mathrm{total}}$ and $v_{yy} = 1 - (1 - \beta)\nu\varepsilon_{\mathrm{total}}$. Third, the positions of the Dirac points shift in momentum space. These shifts are described by gauge vector potentials $\boldsymbol{A_i}$, where the index i = 1, 2, 3 labels the $K$ and $K'$ valleys (Fig.\ 2a). Fourth, the subband positions also move in momentum space as the nanotubes' circumferences shrinks under strain (Fig.\ 2a). There is no strain in the source/drain SWCNT contacts, thus the Hamiltonian (Eq.\ \ref{Eq:H_contacts}) has no strain-induced terms. In the contacts, the term $\Delta\mu_{\text{G}}$ is replaced with $\mu_{\text{contact}}$, which is the fixed contact doping due to the charge transfer from the gold film to the tube.

In a strained SWCNT channel, the Dirac equation and the plane wave solution are\cite{McRae24}:

\begin{equation}
\hbar v_{F} \left(\sigma_{x},\sigma_{y}\right).(\boldsymbol{\bar{I}}+(1-\beta)\boldsymbol{\bar{\varepsilon}}).\begin{pmatrix}\tilde{k_{x}}\\\\\tilde{k_{y}}\end{pmatrix}\boldsymbol{\Psi}=\tilde{E_{n}}\boldsymbol{\Psi}
\end{equation}
\\
\begin{myequation}
\boldsymbol{\Psi}_{\tilde{k_y},y,\tilde{k_x},x}=a_N\begin{pmatrix}1\\\mathrm{sgn}(\tilde{k_F})\frac{\tilde{k_x}v_{xx}+i\tilde{k_y}v_{yy}}{\sqrt{(\tilde{k_x}v_{xx})^2+(\tilde{k_y}v_{yy})^2}}\end{pmatrix}e^{i\tilde{k_y}y}e^{i\tilde{k_x}x}
\end{myequation}
\\
where $\tilde{k_x}$, $\tilde{k_y}$ are the $x$ and $y$ component of the wavevector $\tilde{k_F}=\tilde{E_n}/(\hbar v_F)$ in the strained channel, $\text{sgn}(\tilde{k_F}) = \pm 1$ refers to the conduction and valence bands respectively, $a_N$ is a normalization coefficient.

\subsection{Transmission across uniaxial-strain barriers} \label{Subsec: Transmission across uniaxial-strain barriers}
\begin{figure}[!htbp]
\centering
\includegraphics[scale=1.2]{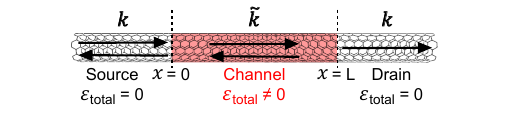}
\caption []{Diagram showing a 1D representation of the reflection and transmission of charge carriers across a strained SWCNT channel and unstrained SWCNT source/drain contacts. The source-channel and channel-drain boundaries are located at $x=0$ and $x=L$, respectively.}
\label{Fig. S1}
\end{figure}

Wavevectors in the contacts and channel are denoted as $\boldsymbol{k}=k_x\hat{x}+k_y\hat{y}$ and $\widetilde{\boldsymbol{k}}=\widetilde{k}_x\hat{x}+\widetilde{k}_y\hat{y}$, respectively, where $k_y = 2\pi N/C_h$, $C_h$ is tube circumference and $N$ is an integer. The wavefunction for the $N$-th mode (subband) in the different regions as shown in Fig.\ \ref{Fig. S1} can then be obtained:

\begin{equation}
\boldsymbol{\Psi}=\begin{cases}\boldsymbol{\Phi}_{S},&\mathrm{if~}x<0\\\\ \boldsymbol{\tilde{\Phi}},&\mathrm{if~}0<x<L\\\\ \boldsymbol{\Phi}_{D},&\mathrm{if~}x>L\\\end{cases}
\end{equation}

\begin{equation}
\boldsymbol{\Phi}_{S}=\boldsymbol{\Psi}_{k_{y},y,k_{x},x}+r_{N}\boldsymbol{\Psi}_{k_{y},y,-k_{x},x}
\end{equation}

\begin{equation}
\boldsymbol{\tilde{\Phi}}=\alpha_{N}\boldsymbol{\Psi}_{\tilde{k_{y}},y,\tilde{k_{x}},x}+\beta_{N}\boldsymbol{\Psi}_{\tilde{k_{y}},y,-\tilde{k_{x}},x}
\end{equation}

\begin{equation}
\boldsymbol{\Phi}_{D}=t_{N}\boldsymbol{\Psi}_{k_{y},y,k_{x},x-L}
\end{equation}
\\
Here, $r_{N}$ and $t_{N}$ represent the reflection and transmission amplitudes, while $\alpha_{N}$ and $\beta_{N}$ are coefficients. Due to the continuity of $\Psi$ at the boundaries $x=0$ (source-channel edge) and $x=L$ (channel-drain edge), we obtain the following boundary conditions.

\begin{equation} \label{Eq: BC_at_0}
\boldsymbol{\Psi}_{k_y,y,k_x,0}+r_N\boldsymbol{\Psi}_{k_y,y,-k_x,0}=\alpha_N\boldsymbol{\Psi}_{\tilde{k_y},y,\tilde{k_x},0}+\beta_N\boldsymbol{\Psi}_{\tilde{k_y},y,-\tilde{k_x},0}
\end{equation}

\begin{equation} \label{Eq: BC_at_L}
\alpha_N\boldsymbol{\Psi}_{\tilde{k_y},y,\tilde{k_x},L}+\beta_N\boldsymbol{\Psi}_{\tilde{k_y},y,-\tilde{k_x},L}=t_N\boldsymbol{\Psi}_{k_y,y,k_x,L-L}
\end{equation}
\\
Note that the coefficient $a_N$ has dropped out. We can then obtain the transmission amplitude of $N$-th mode by solving Eq. \ref{Eq: BC_at_0} and Eq. \ref{Eq: BC_at_L}:

\begin{equation} \label{Eq:t_n_Str}
t_N=\frac{k_x\tilde{k_x}v_{xx}}{k_x\tilde{k_x}v_{xx}\cos\left(\tilde{k_x}L\right)-i\left(-k_y\tilde{k_y}v_{yy}+\mathrm{sgn}(k_F\tilde{k_F})\sqrt{(k_x^2+k_y^2)((\tilde{k_x}v_{xx})^2+(\tilde{k_y}v_{yy})^2)}\right)\sin\left(\tilde{k_x}L\right)}
\end{equation}
\\
where $k_F = \mu_\text{contact}/(\hbar v_F)$ is the magnitude of the Fermi wave vector in the contacts, and $L$ is the channel's length. The transmission probability of the $N$-th mode is given by $|t_N|^2$:

\begin{equation} \label{Eq:T_n_Str}
T_{N}=\frac{(k_x\tilde{k_x}v_{xx})^2}{(k_x\tilde{k_x}v_{xx})^2\cos^2(\tilde{k_x}L)+(k_F\tilde{k_F}-k_y\tilde{k_y}v_{yy})^2\sin^2(\tilde{k_x}L)}
\end{equation}
\\
The only remaining unknown term is the quantized y-momentum $\widetilde{k_y}$, which is discussed in Sec. \ref{Sec: quantized y-momentum}. By summing over all conduction modes, one can find the ballistic conductance in strained SWCNTs:

\begin{equation}
G=\frac{4e^2}{h}\sum_{N=0}^{N_{\text{max}}}T_{N}
\end{equation}
\\
where the number of modes in the summation, $N_{\text{max}}$, is determined by the contact potential $\hbar v_\text{F}k_y \leq \mu_\text{contact}$, and the factor of 4 arises from the spin and valley degeneracies. For quasi-metallic nanotube contacts, the energy difference between two adjacent modes (i.e. subbands) is determined by the diameter due to its periodic boundary condition along the circumference: $\Delta k_y = 2/d$. Because the diameter of any single-wall carbon nanotube is very likely smaller than $2.0$ nm \cite{Yamada06}, the minimum subband energy spacing is $\approx$ 0.6 eV. Because $\mu_{\text{contact}}$ is lower than this value experimentally, there is a single conduction mode ($N_{\text{max}} =$ 0) in our quasi-metallic SWCNT transistors.

\subsection{Quantized y-momentum in the strained SWCNT channel} \label{Sec: quantized y-momentum}
\begin{figure}[!htbp]
\centering
\includegraphics[scale=1.2]{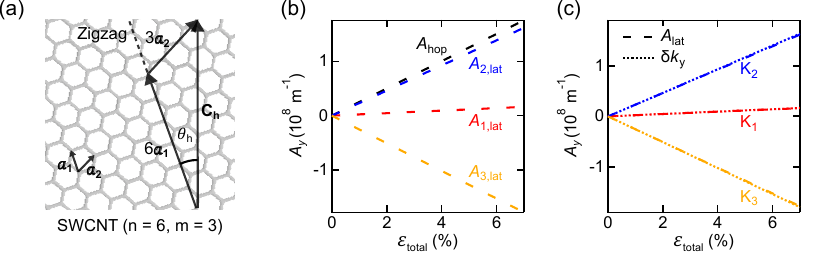}
\caption []{Chirality and strain-induced vector potentials in SWCNTs. (a) A ($n = 6, m = 3$) SWCNT unrolled into a graphene ribbon. The chiral angle $\theta_h = 19^\circ$, chiral vector $\boldsymbol{C}_{\text{h}}$, and lattice vectors are shown. (b) ${A_{\text{i,lat,}y}}$ and ${A_{\text{hop,}y}}$ versus $\varepsilon_{\text{total}}$ for a ($n = 12, m = 9$) tube, with $\theta_h = 25^\circ$. (c) The subband shifts $\delta \tilde{k}_{\text{i,}y}$ (dotted lines) versus $\varepsilon_{\text{total}}$ always exactly cancel the $A_{\text{i,lat},y}$ shifts(dashed lines). The quantities shown are for a (12,9) tube.}
\label{Fig: S2}
\end{figure}

When uniaxial strain is applied to SWCNTs, both the nearest-neighbour carbon atom distances and the hopping amplitudes between atoms are modified. These two types of changes can be described by the gauge potentials $\boldsymbol{A}_{i,\text{lat}}$ and $\boldsymbol{A}_{\text{hop}}$, respectively \cite{Kitt12}. The total vector potential $\boldsymbol{A}_{\text{i}} = \boldsymbol{A}_{\text{i,lat}} + \boldsymbol{A}_{\text{hop}}$ indicates the shift of the Dirac points under strain in $K_{\text{i}}$ valley, where $i = 1, 2, 3$. By symmetry, $\bm{A'_{\text{i}}}$ in the $K'_{\text{i}}$ valleys are equal to $\bm{-A_{\text{i}}}$. The $\bm{A_{\text{i}}}$ are given by Eq. \ref{Eq:A}:

\begin{subequations} \label{Eq:A}
\begin{equation}
\begin{aligned}
\boldsymbol{A}_{\text{hop}}& =\frac{\beta\varepsilon(1+\nu)}{2a}\begin{pmatrix}\sin3\theta_h\\\\\cos3\theta_h\end{pmatrix}  \\
\end{aligned}
\end{equation}
\\
\begin{equation}
\begin{aligned}
\boldsymbol{A}_{1,\text{lat}}& =\frac{2\pi\varepsilon}{3a}\begin{pmatrix}-\cos\theta_h - \frac{1}{\sqrt{3}}\sin\theta_h\\\\\frac{1}{\sqrt{3}}\nu\cos\theta_h -\nu\sin\theta_h\end{pmatrix}
\end{aligned}
\end{equation}
\\
\begin{equation}
\begin{aligned}
\boldsymbol{A}_{2,\text{lat}}& =\frac{2\pi\varepsilon}{3a}\begin{pmatrix}\cos\theta_h-\frac{1}{\sqrt{3}}\sin\theta_h\\\\\frac{1}{\sqrt{3}}\nu\cos\theta_h + \nu\sin\theta_h\end{pmatrix}
\end{aligned}
\end{equation}
\\
\begin{equation}
\begin{aligned}
\boldsymbol{A}_{3,\text{lat}}& =\frac{4\pi\varepsilon}{3\sqrt{3}a}\begin{pmatrix}\sin\theta_h\\\\-\nu\cos\theta_h\end{pmatrix}
\end{aligned}
\end{equation}
\end{subequations}
Here, $a=1.42 \text{Å}$ is the nearest neighbor carbon-carbon distance under zero strain. $\theta_h$ is the chiral angle defined by the chiral vector $\boldsymbol{C}_{\text{h}} = n\boldsymbol{a}_1 + m\boldsymbol{a}_2$ with respect to the zigzag direction, where $\boldsymbol{a}_1$ and $\boldsymbol{a}_2$ are the unit vectors of the hexagonal carbon lattice (Fig.\ \ref{Fig: S2}a) \cite{Charlier07}. Note that the crystal orientation notation $\theta_h$ we use here differs from a similar quantity used in graphene devices\cite{McRae24}, where a crystal angle $\theta$ was defined between the device's $x$-axis and zigzag direction. Fig.\ \ref{Fig: S2}b shows how $\varepsilon_{\text{total}}$ modifies linearly the $y$-components of the vector potentials, ${A_{\text{i,}y}}$, in a ($n=12,m=9$) SWCNT where $\theta_h = $25$\degree$.

While applying a uniaxial strain along the tube axis, the circumference of the tube decreases by $\varepsilon_{yy} = -\nu \varepsilon_{\mathrm{total}}$. Thus, the periodic boundary condition along the circumference becomes:

\begin{equation}
\tilde{k_y}(1+\varepsilon_{yy})C_h=2\pi N
\end{equation}
\\
where $N$ is an integer. As shown in the main text Fig.\ 2a, when there is no strain, the subbands (black dashed lines) pass through the Dirac points in quasi-metallic nanotubes. Meaning that there is a zero $y$-momentum $\tilde{k_y}=0$ component in the allowed subband. When strain is applied, the subbands (red dashed lines) move as shown in Fig.\ 2a and $\tilde{k_y}$ in $K_i$ valley is determined both by the Dirac points shift $A_{i,y}$ and the subbands shift $\delta \tilde{k}_{i,y}$. For example, in a quasi-metallic SWCNT ($n=12, m=9$) under $4\%$ strain, $A_{1,y} =$ 1.1 $\times$ 10$^{8}$ m$^{-1}$, $A_{2,y} =$ 2.0 $\times$ 10$^{8}$ m$^{-1}$, $A_{3,y} =$ -0.02 $\times$ 10$^{8}$ m$^{-1}$, while  $\delta \tilde{k}_{1,y} =$ 0.1 $\times$ 10$^{8}$ m$^{-1}$, $\delta \tilde{k}_{2,y} =$ 1.0 $\times$ 10$^{8}$ m$^{-1}$, $\delta \tilde{k}_{3,y} =$ -1.02 $\times$ 10$^{8}$ m$^{-1}$. The strain-induced change in $\tilde{k_y}$ is given by $\delta \tilde{k}_{i,y} - A_{i,y}$ which gives -1.0 $\times$ 10$^{8}$ m$^{-1}$ for all $K_i$ valleys. Fig.\ \ref{Fig: S2}c shows that $\delta \tilde{k}_{i,y}$ is always equals to the lattice-induced term $A_{\text{i,lat,}y}$ under different strain. Thus, in general,

\begin{equation} \label{Eq:ky_shift}
\tilde{k}_y = k_y - A_{i,y} + \delta \tilde{k}_{i,y} = k_y - (A_{\text{i,lat},y} + A_{\text{hop},y}) + A_{\text{i,lat},y} = k_y - A_{\text{hop},y}
\end{equation}
We can use Eq. \ref{Eq:ky_shift} to simplify Eq. \ref{Eq:T_n_Str}, and obtain:

\begin{equation} \label{Eq:T_n_Str_A}
T_{\xi}=\frac{(k_x\tilde{k_x}v_{xx})^2}{(k_x\tilde{k_x}v_{xx})^2\cos^2(\tilde{k_x}L)+(k_F\tilde{k_F}-k_y(k_y-\xi A_{\text{hop},y})v_{yy})^2\sin^2(\tilde{k_x}L)}
\end{equation}
\\
where $\xi =$ 1 (-1) in the $K_i$ ($K_i'$) valley,  $\tilde{k_F} = |\bm{\tilde{k}}|$, $k_F = |\bm{k}|$, $k_x = \sqrt{{k_F}^2-{k_y}^2}$, $\tilde{k_x}={v_{xx}}^{-1}\sqrt{\tilde{k_F}^2-{v_{yy}}^2\tilde{k_y}^2}={v_{xx}}^{-1}\sqrt{\tilde{k_F}^2-{v_{yy}}^2({k_y - \xi A_{\text{hop},y}})^2}$. By summing over all valleys, the conductance of the nanotube transistor is finally obtained as
\begin{equation} \label{Eq:G_total}
G=\frac{4e^2}{h}\frac{1}{2}\sum_\xi T_{\xi}
\end{equation}

\section{Density of states and quantum capacitance effects in SWCNT transistors}
From Eq. \ref{Eq:T_n_Str_A} and Eq. \ref{Eq:G_total}, we can calculate the conductance of a strained SWCNT device at any gate-induced energy shift $\Delta \mu_\text{G}$. However, in experiments, it is rather a gate voltage $V_\text{G}$ that is applied to the device. In this section, we discuss how to convert $\Delta \mu_\text{G}$ to $V_\text{G}$.

In our proposed SWCNT devices, the channel and the gate (silicon substrate) form a capacitor. Its capacitance is determined by both the usual geometric capacitance $C_\mathrm{G}$ and a quantum capacitance $C_\mathrm{DOS}$ stemming from the density of states in the SWCNT \cite{Ilani06}:

\begin{equation}
C_\mathrm{total}^{-1}=C_\mathrm{DOS}^{-1}+C_\mathrm{G}^{-1}
\end{equation}
where $C_\mathrm{G}$ is given by Eq. \ref{Eq:C_G} above, and $C_\mathrm{DOS}$ is:

\begin{equation}
C_\mathrm{DOS}=e^2 g(\mu_\text{channel})
\end{equation}
where $g(\mu_\text{channel})$ is the density of states at the chemical potential $\mu_\text{channel}= \Delta\mu_\text{G} + e\phi_{\varepsilon}$. As noted above, a band gap in strain-induced in quasi-metallic SWCNT, and its value is simply: $E_\text{gap} = 2\hbar v_{F} |A_{\text{hop},y}|$. Then, $g(\mu_\text{channel})$ can be written as:

\begin{equation}
\begin{aligned}
g(\mu_\text{channel})&=\frac{4}{\pi\hbar v_\mathrm{F}}\big[1-(E_\mathrm{gap}/2\mu_\text{channel})^2\big]^{-1/2}\\
&=\frac{4}{\pi\hbar v_\mathrm{F}}\big[1-(\hbar v_{F} |A_{\text{hop},y}|/\mu_\text{channel})^2\big]^{-1/2}
\end{aligned}
\end{equation}

The above capacitances determine the conversion ratio between the gate voltage and chemical potential \cite{Ilani06}:

\begin{equation}
\frac{\mathrm{d}V_{\mathrm{G}}}{\mathrm{d}\mu_{\text{channel}}}=\frac{C_{\mathrm{DOS}}}{eC_{\mathrm{tot}}}
\end{equation}
\\
One can then find the gate voltage $V_{\mathrm{G}}$ from the gate energy $\Delta\mu_\text{G}$:

\begin{equation}
V_{\mathrm{G}} = \frac{1}{e}\int_{0}^{\Delta\mu_{\text{G}}}\frac{C_{\mathrm{DOS}}}{C_{\mathrm{tot}}}\text{d}\Delta\mu_{\text{G}}
\end{equation}
\\
Note that because $C_{\mathrm{DOS}}$ is strain-dependent, the $V_{\mathrm{G}}$ giving a specific $\Delta\mu_\text{G}$ is not the same at different strains.

\section{Conductance data for quasi-metallic SWCNTs with different tube diameters}
Figure 4 of the main text shows the mechanical control of the quantum phase of current in a (12,9) SWCNT channel whose diameter $d \approx 1.5$ nm. To prove that this prediction is valid for a broad range of quasi-metallic SWCNTs, we present below additional data for quasi-metallic tubes with different diameters and chiralities.

First, we look at a SWCNT with a similar diameter $d \approx 1.5$ nm, but a different chirality (14,8). Figure \ref{Fig: Chirality}a is a $G$ - $\Delta\mu_{\text{G}}$ - $\varepsilon_{\text{total}}$ color map for a (14,8) quasi-metallic tube. Comparing these data to the (12,9) data in Fig.\ 4(a), we notice that the blue region (bandgap) is wider, which indicates a larger $A_\text{hop,y}$ in the (14,8) tube at the same strain. A clear pattern of interferences is observed in all cases. Fig. \ref{Fig: Chirality}b shows the $G$ data extracted along the grey and black cuts in Fig.\ref{Fig: Chirality}a, where $\Delta\mu_\text{G} = 0.1$ eV and -0.2 eV, respectively. We note that changing $\varepsilon_{\text{total}}$ from 3.5 \% to 3.7 \% suffices to change by $\pi$ the phase of the $G$ interference. Moreover, the amplitude of the interference is large and reaches up to $\Delta G / G_{\text{max}} \approx$ 0.8.

\begin{figure}[!htbp] \label{Fig: Chirality}
\centering
\includegraphics[scale=1.2]{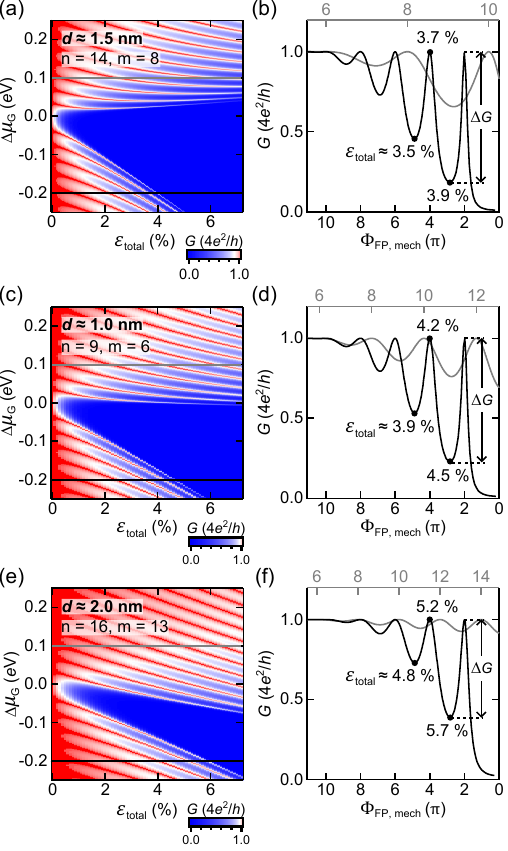}
\caption []{Transport data for strain transistors based on quasi-metallic SWCNTs of different diameters and chiralities. For a tube (14,8) with $d\approx$1.5 nm, (a) $G$ - $\Delta\mu_{\text{G}}$ - $\varepsilon_{\text{total}}$ data and (b) $G$ - $\Phi_{\text{FP, mech}}$ data extracted from (a) along the grey and black cuts.  For a tube (9,6) with $d\approx$1.0 nm, (c) $G$ - $\Delta\mu_{\text{G}}$ - $\varepsilon_{\text{total}}$ data and (d) $G$ - $\Phi_{\text{FP, mech}}$ data extracted from (c) along the grey and black cuts. For a tube (16,13) with $d\approx$ 2.0 nm, (e) $G$ - $\Delta\mu_{\text{G}}$ - $\varepsilon_{\text{total}}$ data and (f) $G$ - $\varphi_{\text{FP, mech}}$ data extracted from (c) along the grey and black cuts.}
\label{Fig: Chirality}
\end{figure}

Next, we look at two SWCNTs with different diameters: $n = 9, m = 6$, $d \approx 1.0$ nm and $n = 16, m = 13$, $d \approx 2.0$ nm. There are clear interference patterns both in Fig. \ref{Fig: Chirality}c and Fig. \ref{Fig: Chirality}e. The latter has a smaller bandgap region. Figures \ref{Fig: Chirality}d,f confirm that using the available $\varepsilon_{\text{total}}$ range one can fully tune the $G$ interferences in these tubes.

\section{Main sources of uncertainty in the conductance calculations}
In our model, we incorporated three experimental parameters whose values were taken from literature: the contact region Fermi energy $\mu_\text{contact}$, the scalar energy pre-factor $g_\varepsilon$, and the electronic Gr\"{u}neisen parameter $\beta$. Their exact value has some uncertainty in literature and may also vary slightly in different SWCNT devices. In addition, we made two approximations when writing the Hamiltonian in Eq. \ref{Eq:H_channel}, namely we included the tensile strain effects only to first-order and we did not include any shear (torsional) strain. In this section, we discuss the magnitude of the uncertainty introduced by theses parameters and approximations, and find that they do not introduce uncertainties on $G$ beyond about 20$\%$. The model is therefore sound for describing realistic experiments which typically have other experimental sources of uncertainty of a similar magnitude (e.g. the uncertainty on $L$ can be around $\pm$ 10 nm \cite{McRae17, McRae24}, which amounts to 20 $\%$ of $L=$ 50 nm).

\subsection{Impact from input parameters $\mu_\textmd{contact}$, $g_\varepsilon$, $\beta$}
As stated in the previous section Sec.\ \ref{Subsec: Transmission across uniaxial-strain barriers}, when $\mu_\text{contact}$ is smaller than the minimum energy of the second subband, there is a single conduction mode in the channel. Here, we calculate the conductance with $\mu_\text{contact} = 0.12$ eV and $0.25$ eV in a SWCNT $n = 12, m = 9$. Both values are smaller than the required energy to reach the second subband, which is $\approx$0.60 eV. Fig.\ \ref{Fig: Input_Parameters}a shows the conductance under different strain at $\Delta\mu_\text{G} = -0.2$ eV. We see clearly that $\mu_\text{contact}$ has no impact on the calculated $G$ data, and thus does not introduce any significant uncertainty.

$g_\varepsilon$ changes the scalar energy $e\phi_\varepsilon = g_\varepsilon (1-\mu) \varepsilon_\text{total}$, which leads to lateral shifts of the gate voltage position of the charge neutrality point. Fig.\ \ref{Fig: Input_Parameters}b shows $G - V_\text{G}$ curves calculated using different $g_\varepsilon$ values. By increasing $g_\varepsilon$ from 2.6 eV to 3.0 eV, which is approximately the uncertainty range, the charge neutrality point shifts from $V_\text{G} \approx -1.8$ V to $\approx -3.1$ V. Most importantly, the impact of uncertainty in $g_\varepsilon$ is a rigid shift of the entire calculate $G$ data spectra and does not change its features. Moreover, the amplitude of the rigid $V_{\text{G}}$ shift is much smaller than the available experimental $V_{\text{G}}$ range. This ensures that the uncertainty on $g_\varepsilon$ does not introduce a meaningful error in connecting the calculated $G$ to experimental measurements.

A reasonable range of uncertainty on $\beta$ is shown in Fig.\ \ref{Fig: Input_Parameters}c, where the black and red data are calculated with $\beta =$ 2.5 and 3.0, respectively. The difference between the two data sets shows that the uncertainty on $\beta$ has only minor effects on the key features of $G$. For instance, the magnitude of its oscillations, the spacing between its maxima, and the ability to fully tune the phase of $G$ mechanically are only slightly modified. Therefore, the calculated data present in the main text will lead to robust experimental features independently of the exact value of $\beta$ .

\begin{figure}[!htbp] \label{Fig: Input_Parameters}
\centering
\includegraphics[scale=1.2]{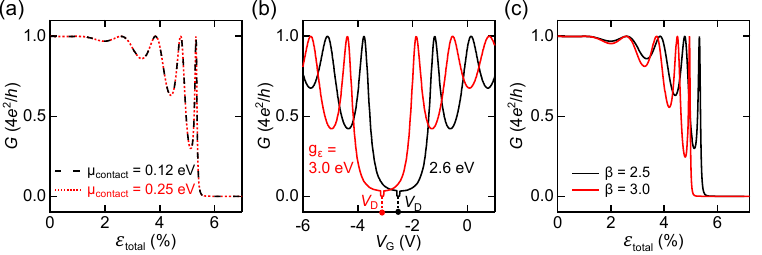}
\caption []{(a) $G$ vs $\varepsilon_\text{total}$ at $\Delta \mu_\text{G} = -0.2$ eV in a quasi-metallic nanotube $n = 12, m = 9$. The dashed-black data and dotted-red data are calculated with $\mu_\text{contact} = 0.12$ eV and $0.25$ eV, respectively. (b) $G$ vs $V_\text{G}$ at $\varepsilon_\text{total} = 4$ \% in a quasi-metallic nanotube $n = 12, m = 9$. Black and red solid lines show the calculated data with $g_\varepsilon = 2.6$ eV and 3.0 eV, respectively. (c) $G$ vs $\varepsilon_\text{total}$ at $\Delta \mu_\text{G} = -0.2$ eV in a (12, 9) tube. The black and red data are calculated with $\beta =$ 2.5 and 3.0, respectively.}
\label{Fig: Input_Parameters}
\end{figure}

\subsection{Impact of first-order strain approximation in the Hamiltonian}
The expressions for the $A_{\text{i}}$ in Eq.\ \ref{Eq:A} and the strain tensor $\bar{\varepsilon}$ in Eq.\ \ref{Eq:H_channel} only consider first-order strain corrections and uniaxial strain. This section discusses the impact of second-order strain corrections and/or unintentional torsional strain on $G$.

The wave vector $\boldsymbol{\tilde{k}}$ is modified by the strain-induced vector potentials. The second-order correction to $A_\text{hop,}y$ will affect the propagation angle $\Theta$ and thus the transmission probability across the channel. Using the calculations of \textit{Oliva-Leyva et al.}\cite{Oliva_Leyva17}, the effect of including second order strain corrections in a (14,8) tube changes $A_\text{hop,}y$ from 3.3 $\times$ 10$^{8}$ m$^{-1}$ to 3.6 $\times$ 10$^{8}$ m$^{-1}$ for $\varepsilon_{\text{total}}=$ 7\%. We see that even at the largest strains considered in our work, the inclusion of the second order corrections only represented a 10 $\%$ correction to the calculated $A_{\text{i}}$, and leads to a similar change in the calculated $G$. Given that experimental uncertainties in device dimensions\cite{McRae24} (i.e. nanofabrication) are typically $\geq$ 10 $\%$, the first-order strain approximation still provides an accurate model to guide experiments.

Another source of uncertainty in connecting quantitatively experiments and theory could be if a torsional (shear) strain is present in the fabricated devices. From previous experimental work \cite{Island11, McRae17}, we know that a realistic upper bound for such spurious torsional strain is $\approx 1^\circ$. The dominant effect of such a shear strain in SWCNTs is the tuning of the bandgap. This torsion-induced bandgap is smallest for zigzag tubes and largest for armchair tubes \cite{Yang99}. For the SWCNTs mentioned in our work (quasi-metallic, $d\approx$ 1.5 nm), a $1^\circ$ torsion creates a bandgap of $\approx$ 140 meV. This energy scale is equivalent to the band gap generated by a tensile strain of 2.6\%. The effect of this \textit{maximal} spurious torsion is therefore smaller than the thermally-induced uniaxial strain $\varepsilon_{\text{thermal}}\approx 3.2 \%$. Moreover, any built-in torsion during nanofabrication would remain constant while measuring a well-functioning QTS transistor (Fig.\ 1(a) of the main text). Thus, the only mechanically-tunable strain would remain the uniaxial $\varepsilon_{\text{mech}}$ (Fig.\ 1(b)). The impact of any torsional strain would therefore be equivalent to a rigid, and modest in magnitude, offset of the tunable strain range in the device. It would not lead to any qualitative change in our predictions for experiments: strain tunable bandgaps, strain tunable amplitude of the Fabry-P\'{e}rot interferences, and strain-controlled quantum phase of the transistor's current.

\bibliographystyle{api2}
\linespread{1.0}
\bibliography{supplemental}